\documentclass[aps,prl,twocolumn,superscriptaddress,showpacs]{revtex4-1}
\usepackage{graphicx}
\usepackage{amsmath,amssymb}
\usepackage[dvips]{color}

\newcommand{\vect}[1]{{\ensuremath{\boldsymbol{#1}}}}

\begin{document} 
  
  \title{ Reactivity Boundaries to Separate the Fate of
 a Chemical Reaction Associated with an Index-two saddle}

  \author{Yutaka Nagahata}
  \email{yutaka\_nagahata@mail.sci.hokudai.ac.jp}
  \affiliation{Graduate School of Life Science, Hokkaido University}
  \author{Hiroshi Teramoto}
  \email{teramoto@es.hokudai.ac.jp}
  \affiliation{Graduate School of Life Science, Hokkaido University}
  \affiliation{Molecule and Life Nonlinear Sciences Laboratory, Research Institute for Electronic Science, Hokkaido University, Kita 20 Nishi 10, Kita-ku, Sapporo 001-0020, Japan}
  \author{Chun-Biu Li}
  \email{cbli@es.hokudai.ac.jp}
  \affiliation{Molecule and Life Nonlinear Sciences Laboratory, Research Institute for Electronic Science, Hokkaido University, Kita 20 Nishi 10, Kita-ku, Sapporo 001-0020, Japan}
  \author{Shinnosuke Kawai}
  \email{skawai@es.hokudai.ac.jp}
  \affiliation{Graduate School of Life Science, Hokkaido University}
  \affiliation{Molecule and Life Nonlinear Sciences Laboratory, Research Institute for Electronic Science, Hokkaido University, Kita 20 Nishi 10, Kita-ku, Sapporo 001-0020, Japan}
  \author{Tamiki Komatsuzaki}
  \email{tamiki@es.hokudai.ac.jp}
  \affiliation{Graduate School of Life Science, Hokkaido University}
  \affiliation{Molecule and Life Nonlinear Sciences Laboratory, Research Institute for Electronic Science, Hokkaido University, Kita 20 Nishi 10, Kita-ku, Sapporo 001-0020, Japan}
  
  \date{\today}
  
\begin{abstract}
  Reactivity boundaries that divide the destination and the origin of trajectories are of crucial importance to reveal the mechanism of reactions. 
  We investigate whether such reactivity boundaries can be extracted for higher index saddles in terms of a nonlinear canonical transformation successful for index-one saddles by using a model system with an index-two saddle.
  It is found that the true reactivity boundaries do not coincide with those extracted by the transformation taking into account a nonlinearity in the region of the saddle even for small perturbations, and the discrepancy is more pronounced for the less repulsive direction of the index-two saddle system.
  The present result indicates an importance of the global properties of the phase space to identify the reactivity boundaries, relevant to the question of what reactant and product are in phase space, for saddles with index more than one.
  \end{abstract}
  
  
  \maketitle
  
  
  Saddle points and the dynamics in their vicinities play crucial roles in chemical reactions.
  A saddle point on a multi-dimensional potential energy surface is defined as a stationary point at which the Hessian matrix does not have zero eigenvalues and, at least, one of the eigenvalues is negative. 
  Saddle points are classified by the number of the negative eigenvalues, and a saddle that has $n$ negative eigenvalues is called an \textit{index-$n$ saddle}.
  Especially an index-one saddle on a potential surface has long been
  considered to make bottleneck of reactions \cite{Glasstone1941,Steinfeld1989,Bonnet2010}, the sole unstable direction corresponding to the ``reaction coordinate.''
  This is because index-one saddles are considered to be the lowest
  energy stationary point connecting two potential minima, of
  which one corresponds to the reactant and the other to the product,
  and the system must traverse the index-one saddle from the reactant to the product \cite{Zhang2006,Skodje2000,Shiu2004,Bartsch2005a}.

  To estimate reaction rate constants across the saddles, transition state theory was proposed
  \cite{Glasstone1941,Steinfeld1989,Bonnet2010}, by envisaging the existence of a non-recrossing dividing surface (i.e., transition state (TS)) in the region of index-one saddle.
  Recent studies of nonlinear dynamics in the vicinity of index-one saddles
  have revealed the firm theoretical ground for the robust existence of the no-return TS in the phase space
  \cite{KBAr6I,KB01,Wiggins2001,UzerNonlin02,Bartsch2005a,Li2006,Kawai2010a,Hernandez2010,NFLrev,QNFrev,QTDNF,NFrotSK,NFrotUC,Teramoto2011,Koon2000,Jaffe2002,Gabern2005,Gabern2006}
  (see also books \cite{book_adv05,book_adv11} and references therein).  
  The scope of the dynamical reaction theory based on normal form (NF) theory \cite{LL92}
  , a classical analog of Van Vleck perturbation theory, 
  is not limited to only chemical reactions, but also includes,
  for example, ionization of a hydrogen atom under electromagnetic
  fields \cite{Wiggins2001,UzerNonlin02}, isomerization of clusters
  \cite{KBAr6I,KB01}, orbit designs in solar systems \cite{Koon2000,Jaffe2002,Gabern2005,Gabern2006}, and so forth.
  Very recently, these approaches have been generalized to dissipative
  multidimensional Langevin equations
  \cite{Bartsch2005a,Hernandez2010,NFLrev}, laser-controlled chemical
  reactions with quantum effects \cite{QNFrev,QTDNF}, systems with
  rovibrational couplings \cite{NFrotSK,NFrotUC}, and showed the robust
  existence of reaction boundaries even while a no-return TS ceases to
  exist \cite{Kawai2010a}.
 
  For complex molecular systems, the potential energy surface becomes
  more complicated, and transitions from a potential basin to another
  involve not only index-one saddles but also higher index saddles \cite{Minyaev2004,Shida2005,Heidrich1986}. 
  For example, 
  it was shown in a computer simulation of 
  an inert gas cluster containing seven atoms that transitions from
  a solid-like phase to a liquid-like phase occur mostly through 
  index-two saddles rather than through index-one saddle with the increase of
  kinetic temperature \cite{Shida2005}. This indicates that 
  the more rugged a system's energy landscape becomes and/or the more ``temperature''
  increases, the more frequently the system contains higher index
  saddles. 

  To reveal the fundamental mechanism of the passage through a saddle
  with index greater than one, the phase space structure was recently
  studied on the basis of NF theory \cite{Haller2010,Haller2011,Ezra2009a,Collins2011}. 
  For example, the extension of the dynamical reaction theory into
  higher index saddles was discussed
  \cite{Haller2010,Haller2011,Ezra2009a} for a stronger
  repulsive degree of freedom(DoF) \cite{Haller2010,Haller2011}
  and a dividing surface to separate the reactant and the product was proposed for higher index saddles \cite{Collins2011}.
  While these studies are of importance, 
  the stronger repulsive DoF does not necessarily serve as the
  reactive direction, as
  shown for an index-two saddle in structural isomerization of aminoborane \cite{Minyaev2004}. 
  In addition, these studies rely on the
  assumption that NF performed in the region of the saddle can find the
  reactivity boundaries if the perturbation calculation converges
  \cite{Haller2010,Haller2011,Ezra2009a,Collins2011}. 

  In studies of chemical reactions, one needs to assign regions of the phase space as ``reactants" or ``products". 
  Invariant manifolds in the phase space that separate the
  origin and the destination of trajectories have provided us with
  significant implications in the rate calculation and the orbit design
  in non-RRKM systems
  \cite{Koon2000,Jaffe2002,Gabern2005,Gabern2006,KBAr6I,KB01,Wiggins2001,UzerNonlin02,Bartsch2005a,Li2006,Kawai2010a,Hernandez2010,NFLrev,QNFrev,QTDNF,NFrotSK,NFrotUC,Teramoto2011,Koon2000,Jaffe2002,Gabern2005,Gabern2006}
  (see also books \cite{book_adv05,book_adv11} and references therein).  
  In this Letter we investigate how one can identify the reactivity boundaries to determine
  the fate of the reaction for higher index saddles. 
  We analyze a two-DoF Hamiltonian system with an index-two saddle by using NF theory and investigate its applicability in determining if the system undergoes reactions or not.
  We will emphasize the subtlety in defining the ``reactant'' and ``product'' regions in the phase space,
  and point out the difference between the regions defined by NF and those defined by the original coordinates.

\begin{figure}
      \includegraphics[width=8.5cm]{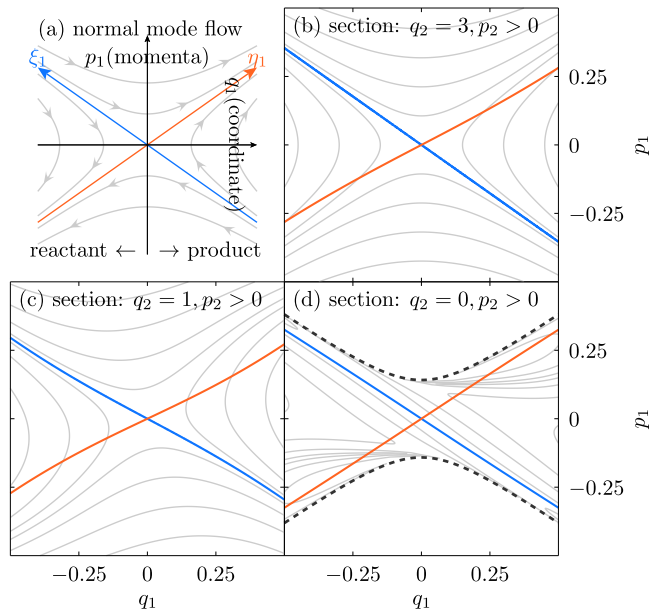}%
      \caption{
        (color online).
        Destination/origin dividing set of trajectories sliced on several sections ($q_2=0,1,3$ with $p_2>0$).
        Each curve represents a set of trajectories (gray, orange, blue), and each initial condition of the set of trajectories is given by a contour of the initial value of the action $I_1$ in the asymptotic region: 
        the initial condition 
        of the destination dividing set of trajectories(blue) is given on that of $q_2=5$ with $p_2>0$ 
        , and that 
        of the origin dividing set of trajectories (orange) is given on the section of $q_2=-5$ with $p_2>0$
        under negative time evolution.
        Here, we have energetically inaccessible region (dashed lines) because of positive kinetic energy $\sum_{j=1}^n p_j^2/2$.
      }
      \label{fig:trj}
    \end{figure}

  If the total energy of the system is just slightly above a stationary point, the $n$-DoF Hamiltonian $H$ can well be approximated by normal mode Hamiltonian $H_0$
  \begin{equation} \label{eq:nm}
    H(\vect{p},\vect{q})
    \approx 
    H_{0}(\vect{p},\vect{q})
    =
    \sum_{j=1}^n \frac{1}{2}(p_j^2+k_j q_j^2)
  \end{equation}
  with normal mode coordinate 
  \vect{q}=$(q_1,\dots,q_n)$ 
  and its conjugate momenta
  \vect{p}=$(p_1,\dots,p_n)$, 
  where $k_j \in \mathbb{R}$ is ``spring constant'' or the curvature of the potential energy surface along the $j$th direction. 
  The constants $k_j$ can be positive or negative.
  If negative, the potential energy is maximum along the $j$th direction.
  Then the direction exhibits an unstable motion corresponding to ``sliding down the barrier,'' and can be regarded as ``reaction coordinate.''
  The index of the saddle corresponds to the number of negative $k_j$.
  Flow of the DoF with negative $k_j$ is depicted in Fig.~\ref{fig:trj}(a). 
  Here one can introduce the following coordinates
  \begin{eqnarray} \label{eq:xieta_nm}
    \eta_j=& (p_j+\lambda_j q_j)/(\lambda_j\sqrt{2})
    , ~~~
    \xi_j =&(p_j-\lambda_j q_j)/\sqrt{2}
    ,
  \end{eqnarray}
  where $\lambda_j=\sqrt{-k_j}$.
  When Eq.\;(\ref{eq:nm}) holds, the action variable defined by $I_j=\xi_j\eta_j$ is an integral of motion,
  and trajectories run along the hyperbolas given by $I_j=const.$ shown by gray lines in Fig. \ref{fig:trj}(a).
  The $\eta_j$- and $\xi_j$-axes run along the asymptotic lines of the hyperbolas in Fig.~\ref{fig:trj}(a).
  One can tell the destination and origin regions of trajectories from the signs of $\eta_j,\xi_j$ as follows:
  If $\eta_j>0$, the trajectory goes into $q_j>0$ and if $\eta_j<0$, then the trajectory goes into $q_j<0$.
  Therefore one can determine the destination of trajectories from the sign of $\eta_j$.
  Similarly, the origin of trajectories can be determined from the sign of $\xi_j$.
  Hereafter we call the set $\eta_j=0$ ``destination-dividing set,'' $\xi_j=0$ ``origin-dividing set,''
  and each of these sets constitute ``reactivity boundaries.''

  The Hamiltonian of Eq.~(\ref{eq:nm}) corresponds to the lowest order (quadratic) part of the Taylor expansion of $H$.
  As total energy of the system increases, one needs to consider higher order terms $H_{\varepsilon}(\vect{p},\vect{q})$:
  \begin{equation} \label{eq:fullH}
    H(\vect{p},\vect{q})
    =
    H_{0}(\vect{p},\vect{q})
    +
    H_{\varepsilon}(\vect{p},\vect{q}),
  \end{equation}
  where $H_{\varepsilon}$ is power series starting from cubic and higher order terms.
  Note that, in this case, the actions $\vect{I}$ are no longer constants of motion.
  However, previous studies \cite{KBAr6I,KB01,Wiggins2001,UzerNonlin02,QNFrev,QTDNF,NFrotSK,NFrotUC} 
  showed that a nonlinear canonical transformation $(p_1,\dots,p_n,q_1,\dots,q_n)\rightarrow(\bar p_1,\dots,\bar p_n,\bar q_1,\dots,\bar q_n)$
  can provide new action variables as constants of motion,
  and the associated degrees of freedom are decoupled with each other (to a certain order of approximation) in the new coordinates.
  Here the new actions and the coordinates are defined in parallel with Eq.~(\ref{eq:xieta_nm}) by using the newly introduced coordinates $(\bar{\vect{p}},\bar{\vect{q}})$:
  \begin{eqnarray} \label{eq:Inf}
    \bar{I}_j = & \bar{\xi}_j \bar{\eta}_j,\cr
    \bar\eta_j = & (\bar p_j+\lambda_j \bar q_j)/(\lambda_j\sqrt{2})
    ,~~~
    \bar\xi_j = & (\bar p_j-\lambda_j \bar q_j)/(\sqrt{2})
    .
  \end{eqnarray}
  The newly introduced coordinates $(\bar{\vect{p}},\bar{\vect{q}})$ are called NF coordinates. 
  The new actions $\bar{I}_j$ are now constants of motion, and consequently the flow around the stationary point follows the contour lines shown in Fig.~ \ref{fig:trj}(a),
  if the  axes are changed to the new coordinate $\bar{q}_1$ and $\bar{p}_1$. 
  Thus one can still know the destination and the origin of trajectories from the signs of $\bar{\eta}$ and $\bar{\xi}$. 
  Note that 
  the NF theory is based on the assumption that linear terms dominate
  dynamics around the saddle and are weakly perturbed by nonlinear terms. 
  Under this assumption $\lambda$s of the normal modes around a
  saddle point dominate the dynamics
  and can extract the integrals of motion if the perturbation
  calculation converges.

  In order to understand whether such reactivity boundaries extracted by NF actually coincide with the true reactivity boundaries
  that determine the asymptotic behavior of a chemical reaction through index-two saddle,
  we scrutinize a two DoF model system with an index-two saddle whose higher order term in Eq. (\ref{eq:fullH}) is
  \begin{equation} \label{eq:nonlinear}
    H_\varepsilon(\vect{p},\vect{q})
    = \varepsilon
    q_1^2 q_2^2 \exp(2-q_1^2-q_2^2).
  \end{equation}
  This nonlinear term is effective locally around $|q_1|=|q_2|=1$, and vanishes in the asymptotic region($|q_1|$ or $|q_2|=\infty$) and in the vicinity of the saddle ($|q_1|$ and $|q_2|\approx 0$).
  In what follows, we employ the system parameters as $E=10^{-2},~\varepsilon=10^{-1},~\lambda_1=1/\sqrt{2}$ and $\lambda_1:\lambda_2=1:\gamma$ (golden ratio).

  In order to observe the trajectories and the destination- and the origin-dividing sets, we take a set of sections of the phase space at some values of $q_j$ ($j=1$ or $2$).
  For example, Fig.~\ref{fig:trj}(d) shows the section at $q_2=0$ with $p_2>0$. 
  There the gray curves are contour lines of the initial value of the normal mode action $I_j$.
  The blue and orange curves are the destination-dividing set and the origin-dividing set, respectively.
  Numerical extraction of the destination-dividing set is carried out as follows:
  First we take a set of points on the line $\eta_1=0$ on the section $q_2=5$ and $p_2>0$.
  This set divides the destination of the trajectories correctly, the large negative values of $q_2$ and the negative sign of $p_2$ ensure that the trajectories will go into the asymptotic region with negative $q_2$, where the flows of the trajectories are given by the normal mode Hamiltonian [see Eqs.~(\ref{eq:fullH}) and (\ref{eq:nonlinear})], and $H_\varepsilon$ becomes negligible for large $|\vect{q}|$ as shown in Fig.~\ref{fig:trj}(a),(b).
  The set is then numerically propagated backward in time into the inner region (smaller values of $q_2$) where the nonlinear term is significant as shown in Fig.~\ref{fig:trj}(c),(d).
  Similarly, the origin-dividing set is calculated by taking a set of points on $\xi_1=0$ on the section of $q_2=-5$ and $p_2>0$, and propagating them forward in time.

\begin{figure}
    \includegraphics[width=8.5cm]{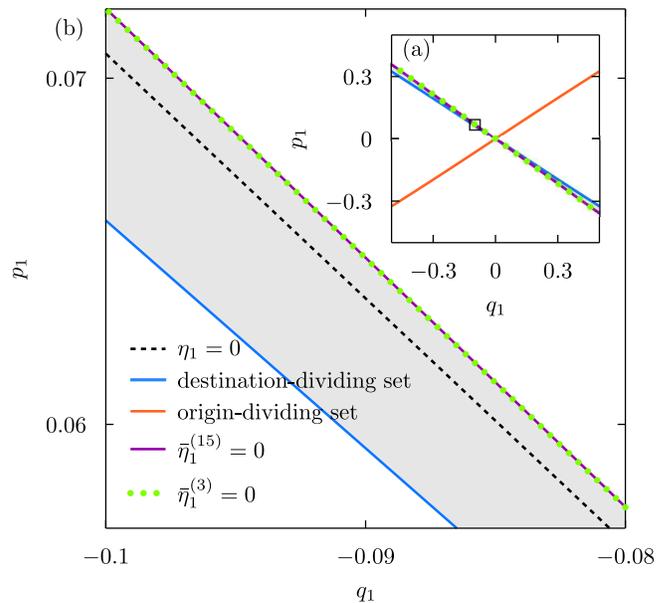}
    \caption{
      (color online).
      Discrepancy between the numerically extracted destination-dividing set and that of NF $\bar{\eta}_1=0$ on the section of $q_2=0$ with $p_2>0$.
      Square depicted in (a) denotes the region which are magnified in (b).
      The shaded areas denote a discrepancy region where the two destination-dividing sets were different.
    }
    \label{fig:dfm}
  \end{figure}

\begin{figure}
    \includegraphics[width=8.5cm]{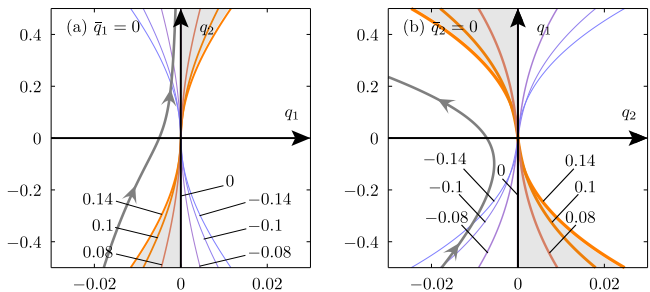}%
    \caption{
      (color online).    
      Contour lines of $\bar{q}_1(\mathbf{p},\mathbf{q}
      |H=E)=0$ and
      $\bar{q}_2(\mathbf{p},\mathbf{q}| H=E)=0$ on
      the $q_1$-$q_2$ space (a) and the $q_2$-$q_1$ space (b) with some fixed values
      of $p_1$ and $p_2$ whose values are indicated in the insets. The gray bold curves denote representative
      trajectories.
      For instance, the discrepancy regions of $\mathrm{sgn}~ q_i \ne \mathrm{sgn}~ \bar{q}_i$ with $p_i=0.14 (i=1,2)$ are denoted by the gray colored areas.
    }
       \label{fig:cnt}
  \end{figure}

  Now we compare the numerically calculated destination- and origin-dividing sets with those calculated by the NF theory.
  Fig.~\ref{fig:dfm} shows the destination-dividing set on the section of $q_2=0$ with $p_2>0$.
  We observe discrepancy between the numerically calculated set and those of NF ($\bar{\eta}_1^{(3)}=0$ and  $\bar{\eta}_1^{(15)}=0$, where the upper indices denote the polynomial order of NF).
  When compared with the normal mode approximation ($\eta_1=0$), 
  it is seen that the effect of the nonlinearity is evaluated in the opposite way in the NF compared to the true destination-dividing set.
  The failure of the NF observed here in calculating the destination-dividing set is not due to the lack of convergence in the perturbation expansion used in the NF theory, because, firstly, the results of the third- and the fifteenth-order of expansion compared in Fig.~\ref{fig:dfm} confirm a good convergence of the NF, and secondly, we have confirmed that the numerically computed trajectories follow the NF destination-dividing set $\bar{\eta}_1=0$ in the saddle region, that is, the set $\bar{\eta}_1=0$ is truly an invariant set.
  Thus the NF describes correctly the dynamics of this system, and the sign of $\bar{\eta}_1$ predicts the destination of the trajectory in the $(\bar{q}_1,\bar{p}_1)$-space.
  However, the ``destination'' predicted from the sign of the NF coordinate $\bar{\eta}_1$ rather refers to the sign of $\bar{q}_1$ in the future, as can be seen from the discussion in Fig.~\ref{fig:trj}(a).
  The NF can fail to predict the destination of trajectories when the sign of $\bar{q}_1$ is different from the originally used position coordinate $q_1$. 
  Figure~\ref{fig:cnt} presents some contour lines of $\bar{q}_1(\mathbf{p},\mathbf{q}|E)=0$ and $\bar{q}_2(\mathbf{p},\mathbf{q}|E)=0$ on the $q_1$-$q_2$ space and the $q_2$-$q_1$ space, respectively,  with some fixed values of $p_1$ and $p_2$. 
  The right (left) hand side region of each contour line in the spaces corresponds to a region of $\bar{q}_j>0$ ($\bar{q}_j<0$) for fixed $p_j$ ($j=1$ in Fig.~\ref{fig:cnt}(a), $j=2$ in Fig.~\ref{fig:cnt}(b)). 
  These plots indicate that there exist regions where the signs of $\bar{q}_j$ and $q_j$ are different,
  and the size of the discrepancy regions ($\mathrm{sgn}~ q_j \neq \mathrm{sgn}~ \bar{q}_j$) tends to enlarge with the increase of $|p_j|$(e.g., see the shaded areas in Fig.~\ref{fig:cnt}). 
  Likewise, such failure of the NF also occurs for $\bar{\xi}_1=0$ in determining the origin.
  As Fig.~\ref{fig:cnt} indicates, such a discrepancy can also occur for $\bar{\eta}_2=0$ and $\bar{\xi }_2=0$.
  Note however that the significance of discrepancy in the NF reactivity boundary is different depending on the instability of these reactive DoFs. 
  Trajectories, denoted by the gray bold curves in Fig.~\ref{fig:cnt}(a)(b), are more strongly repelled along the $q_2$ direction than $q_1$ due to the difference of the repulsion ($\lambda_2>\lambda_1$). 
  The discrepancy between those NF reactivity boundaries and the corresponding destination- and origin-dividing sets is more pronounced along $q_1$ than along $q_2$. 
  It is because trajectories more often enter into the discrepancy region of $\mathrm{sgn}~ q_1 \neq \mathrm{sgn}~ \bar{q}_1$ than that of $\mathrm{sgn}~ q_2 \neq \mathrm{sgn}~ \bar{q}_2$ due to the difference of the repulsion.
  Because we interpret this result in terms of the relative magnitudes of the $\lambda$s without referring to any specific properties of our model,
  similar results are expected to be found generally in the dynamics around index-two saddles in reacting systems
  when linear terms dominate dynamics around the saddle and are weakly perturbed by nonlinear terms.

  In conclusion, we have numerically constructed the destination- and the origin-dividing sets in a two DoF system with an index-two saddle, and compared the results of NF theory with them.
  We have found the failure of the NF in identifying the reactivity boundaries 
  especially along the less repulsive DoF even while the perturbation calculation converges. 
  On the contrary, significant discrepancy was not observed along the strong repulsive DoF, 
  which agrees with the studies \cite{Haller2010,Haller2011}.
  Such discrepancy could also occur in index-one saddles, although the
  difference between $\bar q$ and $q$ have not been found with
  significance in index-one saddles \cite{KBAr6I,KB01,Wiggins2001,UzerNonlin02,Li2006,Kawai2010a,QNFrev,QTDNF,NFrotSK,NFrotUC,Teramoto2011}. 
  This is probably because, in the case of index-one saddles, there is only one repulsive DoF and all the other DoFs are bound 
  so that trajectories have less possibility to go into the discrepancy regions after leaving the region of the saddle.
  
  In the context of studying dynamics of chemical reaction systems, one needs to divide the asymptotic region of the phase space into ``reactants'' and ``products.''
  For the case of the index-one saddle, this division has seemed trivial 
  because we have only one reactive direction (say $q_1$) and therefore only two asymptotic regions ($q_1\rightarrow+\infty$ and $q_1\rightarrow-\infty$).
  For the case of the higher index saddle, however, we have more than one ``reactive'' direction and the division of the phase space is not trivial any more.
  In this study, to define states we designed a model system that becomes separable in the asymptotic region.
  In a general case, the asymptotic ``reactant'' and ``product'' regions must be assigned by referring to the chemical nature of each specific system such as breaking and formation of chemical bonds.
  According to the present results, however such assignment can be different from those made by NF, especially for less repulsive DoF.
  Such less repulsive DoF can sometimes serve as the reactive
  coordinate in molecular systems \cite{Minyaev2004}.
  This indicates that chemical reactions through higher index saddles can involve much richer structures which require reconsideration of the concepts of ``reactant'' and ``product'' themselves.
  Future works, therefore, will need either to modify the NF reaction theory
  to remedy the discrepancy between $q$ and $\bar{q}$, or 
  resort to numerical calculations, although the latter is difficult for high DoF systems.

  We acknowledge Dr. Yusuke Ohtani and Prof. Mikito Toda for their fruitful discussions.
  This work has been partially supported by the Japan Society for the Promotion of Science. 
  The computations were partially performed using the Research Center for Computational Science, Okazaki, Japan.

\bibliography{prl_nagahata}
\end{document}